\documentclass[doublecol]{epl2}

\usepackage[utf8]{inputenc}
\usepackage{graphicx}
\usepackage{dcolumn}
\usepackage{bm}
\usepackage{amsmath}
\usepackage{amssymb}

\usepackage{lipsum}
\usepackage{mathtools}
\usepackage{cuted}

\usepackage{subcaption}
\graphicspath{{img/},{fig/}}

\newcommand{\sech}{\operatorname{sech}}

\newcommand\egaldef{\stackrel{\mbox{\upshape\tiny def}}{=}}

\newcommand\1{\leavevmode\hbox{\rm \small1\kern-0.35em\normalsize1}}

\newcommand\EE{\mathsf{E}}
\def\DD{\displaystyle}

\begin{document}

\title{Spectral Dynamics of Learning Restricted Boltzmann Machines}
\author{A. Decelle\inst{1}\thanks{E-mail: \email{aurelien.decelle@lri.fr}} \and G. Fissore\inst{1,2} \and C. Furtlehner\inst{2}}
\shortauthor{A. Decelle \etal}

\institute{                    
\inst{1}LRI, AO team, B\^at 660 Universit\'e Paris Sud, Orsay Cedex 91405\\
\inst{2}Inria Saclay - Tau team, B\^at 660 Universit\'e Paris Sud, Orsay Cedex 91405
}



\abstract{
The Restricted Boltzmann Machine (RBM), an important tool used in machine learning in particular for unsupervized learning 
tasks, is investigated from the perspective of its spectral properties. Starting from empirical observations, we propose a generic 
statistical ensemble for the weight matrix of the RBM and characterize its mean evolution.
This let us show how in the linear regime, in which the RBM is found to operate at the beginning of the training, the statistical properties of the data drive the selection of the unstable modes of the weight matrix. A set of equations characterizing the non-linear regime is then derived, unveiling in some way how the selected modes interact in later stages of the learning procedure and defining a deterministic learning curve for the RBM.
}


\maketitle

\section{Introduction}
A Restricted Boltzmann machine (RBM)~\cite{Smolensky} constitutes nowadays  a common tool on the shelf of machine learning practitioners. It is a generative model, 
in the sense that it defines a probability distribution, which can be learned to approximate any distribution of data points living in some 
$N$-dimensional space, with $N$ potentially large. It also often constitutes a building block of more complex neural network models~\cite{HiSa,salakhutdinov2009deep}. 
The standard learning procedure called contrastive divergence~\cite{Hinton_CD}  is well documented~\cite{Hinton_guide} although 
being still a not so well understood fine empirical art, 
with many hyperparameters to tune without much guidelines.
At the same time an RBM can be regarded as a statistical physics model, being defined as a Boltzmann distribution
with pairwise interactions on a bipartite graph. Similar models have been already the subject of many studies in the 
80's~\cite{Hopfield,AmGuSo3,Gardner,Derrida-Gardner} which mainly concentrated on the learning capacity, i.e. the number of independent patterns that could be stored in such a model.
The second life of neural networks has renewed the interest of statistical physicists for such models.
Recent works actually propose to exploit its statistical physics formulation to define mean-field based learning methods using TAP 
equations~\cite{TAP_train,huang2015advanced,takahashi2016mean}.
Meanwhile some analysis of its static properties, assuming a given learned weight matrix $W$, have been proposed~\cite{HHuang,Barra} in order 
to understand collective phenomena in the latent representation~\cite{TuMo}, i.e. the way latent variables organize themselves to represent actual data. One common
assumption made in these works is that the weights of $W$ are i.i.d. which as we shall see is unrealistic. 
Concerning the learning procedure of neural networks, many recent statistical physics based analysis have been proposed, 
most of them within teacher-student setting~\cite{ZdKr} which imposes a strong assumption on the data, namely that these are generated from 
a model belonging to the parametric family of interest, hiding as a consequence the role played by the data themselves in the procedure. 
From the analysis of related models~\cite{TiBi,Bou-Kamp}, it is already a well established fact that a selection of the most important modes of the singular value 
decomposition (SVD) of the data is performed in the linear case.
In fact in the simpler context of linear feed-forward models the learning dynamics can be fully characterized by means of the 
SVD of the data matrix~\cite{Ganguli}, showing in particular the emergence of each mode by order of importance regarding singular values.

In this work we follow this guideline in the context of a general RBM. We propose to characterize both the learned RBM and the learning process 
itself by the SVD spectrum of the weight matrix in order to isolate the information content of an RBM. This allows us then to write 
a deterministic learning equation leaving aside the fluctuations. This equation is subsequently analyzed 
first in the linear regime to identify the unstable deformation modes of $W$; secondly at equilibrium assuming the learning is converging, in order 
to understand the nature of the non-linear interactions between these modes and how these are determined from the input data.
In the first section we recall the RBM model and associated learning algorithm. In the second section 
we show how this algorithm can be described by a generic learning equation. Then we first analyze the linear regime and thereafter 
we describe what happens with the binary RBM. A set of dynamical parameters is shown to emerge naturally from the SVD decomposition of the weight matrix.
The convergence toward equilibrium is analyzed and illustrated later with actual tests on the MNIST dataset. 

\section{The RBM and associated learning procedure}\label{sec:rbm}
An RBM is a Markov random field with pairwise interactions defined on a bipartite graph formed by two layers of non-interacting variables: 
the visible nodes  and the hidden nodes representing respectively data configurations and latent representations. The former noted $\bm{s} = \{s_i,i=1\ldots N_v\}$ 
correspond to explicit representations of the data while the latter noted $\bm{\sigma} = \{\sigma_j,j=1\ldots N_h\}$ 
are there to build arbitrary dependencies among the visible units. They play the role of an interacting field among visible nodes. 
Usually the nodes are binary-valued (of boolean type or Bernoulli distributed) but gaussian distributions or more broadly arbitrary distributions on real-valued bounded support are also used~\cite{general_RBM}, ultimately making RBMs adapt for more heterogeneous data sets. 
Here to simplify we assume that visible and hidden nodes will be taken as binary variables $s_i,\sigma_j \in \{-1,1\}$ 
(using $\pm 1$ values has the advantage of symmetrizing the equations hence avoiding to deal with ``hidden'' biases on the variables when considering binary $\{0,1\}$ variables).
Like the Hopfield model~\cite{Hopfield} which can actually be cast into an RBM~\cite{BarraBSC12} an energy function is defined for a configuration of nodes

\begin{equation}
    E(\bm{s},\bm{\sigma}) = -\sum_{i,j} s_i w_{ij} \sigma_j - \sum_{i=1}^{N_v} \eta_i s_i - \sum_{j=1}^{N_h} \theta_j \sigma_j
    \label{eq_ener_rbm}
\end{equation}

and this is exploited to define a joint distribution 
between visible and hidden units, namely the Boltzmann distribution

\begin{equation}
    p(\bm{s},\bm{\sigma}) = \frac{e^{-E(\bm{s},\bm{\sigma})}}{Z}
    \label{eq_proba_rbm}
\end{equation}

\noindent where  $W$ is the weight matrix and $\bm{\eta}$ and $\bm{\theta}$ are biases, or external fields on the variables. 
\( \textstyle Z = \sum_{\bm{s},\bm{\sigma}} e^{-E(\bm{s},\bm{\sigma})} \) is the partition function of the system. The joint distribution between visible 
variables is then obtained by summing over hidden ones.
In this context, learning the parameters of the RBM means that, given a dataset of $M$ samples composed of 
$N_v$ variables, we ought to infer values to $W$, $\bm{\eta}$ and $\bm{\theta}$ such that new generated data obtained by sampling this distribution 
should be similar to the input data. The general method to infer the parameters is to maximize the likelihood of the model, where the pdf (\ref{eq_proba_rbm}) has 
first been summed over the hidden variables
\begin{equation}
  \mathcal{L} = \sum_j \log(2 \cosh(\sum_i w_{ij} s_i + \theta_j)) - \log(Z).
\end{equation}
Different methods of learning have been set up and proven to work efficiently, in particular the 
contrastive divergence (CD) algorithm from Hinton~\cite{Hinton_CD} and more recently TAP based learning~\cite{TAP_train}. 
They all correspond to expressing the gradient ascent on the likelihood as 
\begin{equation}\label{eq:cd}
    \Delta w_{ij} = \gamma \left( \langle s_i \sigma_j p(\sigma_j|\bm{s}) \rangle_{\rm Data} - \langle s_i \sigma_j \rangle_{p_{\rm RBM}} \right)
\end{equation}
where $\gamma$ is the learning rate. Similar equations can be derived for the biases. The main problem is the second term on the rhs of~(\ref{eq:cd}) 
which is not tractable, and various methods basically differ in their way of estimating this term (Monte-Carlo chains, mean field, TAP \ldots).
For an efficient learning the first term also has to be approximated
by making use of random mini batches of data at each step.

\section{Deterministic dynamics of the learning}

In order to understand the dynamics of the learning we first project the CD equation~(\ref{eq:cd})
onto the basis defined by the SVD of $W$.
As a generalization of eigenmodes decomposition to rectangular matrices, the SVD for a RBM is given by

\begin{equation}
\mathbf{W} = \mathbf{U \Sigma} \mathbf{V}^T
\end{equation}

where \(\mathbf{U}\) is an orthogonal \(N_v \times N_h\)  matrix whose columns are the left singular vectors \(\mathbf{u}^{\alpha} \), \(\mathbf{V}\) is an orthogonal \(N_h \times N_h\) matrix whose columns are the right singular vectors \( \mathbf{v}^{\alpha} \) and \( \mathbf{\Sigma} \) is a diagonal matrix whose elements are the singular values \(w_{\alpha}\). The separation into left and right singular vectors is due to the rectangular nature of the decomposed matrix, and the similarity with eigenmodes decomposition is revealed by the following SVD equations

\begin{align}
\mathbf{W} \mathbf{v}^{\alpha} & = w_{\alpha} \mathbf{u}^{\alpha} \nonumber \\
\mathbf{W}^T \mathbf{u}^{\alpha} & = w_{\alpha} \mathbf{v}^{\alpha} \nonumber
\end{align}

We consider the usual situation where $N_h<N_v$, which means that the rank of $W$ is at most $N_h$. 
$W(t)$ represents the learned weight matrix at time $t$. Let $\{w_\alpha(t)\in[0,+\infty[\}$, 
$\{u_\alpha(t)\in{\mathbb R}^{N_v}\}$ and $\{v_\alpha(t)\in{\mathbb R}^{N_h}\}$ such that 
the following decomposition $w_{ij}(t) = \sum_{\alpha} u_i^{\alpha}(t) w_\alpha (t) v_j^\alpha (t)$ holds.
Discarding stochastic fluctuations usually inherent to the learning procedure and letting the learning rate $\gamma\to 0$, the continuous version of~(\ref{eq:cd})
can be recast as follows:
\begin{strip}
\newpage
\begin{align}
    \left( \frac{dw}{dt} \right)_{\alpha \beta} &= 
\delta_{\alpha,\beta}\frac{dw_\alpha}{dt}(t)+
(1-\delta_{\alpha,\beta})\Bigl(w_\beta(t)\Omega_{\beta\alpha}^v(t)+w_\alpha(t)\Omega_{\alpha\beta}^h\Bigr)
= \langle s_\alpha \sigma_\beta \rangle_{\rm Data} - \langle s_\alpha \sigma_\beta \rangle_{\rm RBM}\label{eq:cd1} \\
\Omega_{\alpha \beta}^v (t) = -\Omega_{\beta\alpha}^v &\egaldef \frac{d\bm{u}^{\alpha,T}}{dt} \bm{u}^\beta = \frac{-1}{w_\alpha+w_\beta}\left( \frac{dw}{dt} \right)_{\alpha \beta}^{\rm A} + \frac{1}{w_\alpha - w_\beta}\left( \frac{dw}{dt} \right)_{\alpha \beta}^{\rm S} \label{eq:cd2} \\ 
	\Omega_{\alpha \beta}^h (t)= -\Omega_{\beta\alpha}^h &\egaldef \frac{d\bm{v}^{\alpha,T}}{dt} \bm{v}^\beta = \frac{1}{w_\alpha+w_\beta}\left( \frac{dw}{dt} \right)_{\alpha \beta}^{\rm A} + \frac{1}{w_\alpha - w_\beta}\left( \frac{dw}{dt} \right)_{\alpha \beta}^{\rm S} \label{eq:cd3}
\end{align}
\end{strip}
Here everything is expressed in the reference frame defined by singular vectors of $W$.  
$s_\alpha= \sum_i u_i^\alpha s_i$ and $\sigma_\alpha = \sum_j v_j^\alpha \sigma_j$ represent spin configurations in this frame. 
Note that one has to keep track of the original reference frame to be able to evaluate 
the data and RBM average in particular when the basic variables are discrete.
We have introduced the skew-symmetric rotation generators $\Omega_{\alpha\beta}^{v,h}(t)$ of 
the basis vectors induced by the dynamics. These tell us how the data rotate relatively to this frame. 
The superscript S,A indicate the symmetric (resp. anti-symmetric) part of the matrix.
Note that these equations become singular when some degeneracy occurs in $W$ because then the SVD is not uniquely defined.
This is not really a problem since we are interested in rotations among non-degenerate modes, the rest corresponding to gauge degrees of freedom.
Similar equations can be derived for the fields $\eta_\alpha(t)\egaldef\sum_i\eta_i(t) u_i^\alpha(t)$ and $\theta_\alpha(t)\egaldef\sum_j v_j^\alpha(t)\theta_j(t)$ 
projected onto the SVD modes. At this point we make the assumption that the learning dynamics is 
represented by a trajectory of $(\{w_\alpha(t),\eta_\alpha(t),\theta_\alpha(t),\Omega_{\alpha\beta}^{v,h}(t)\}$, while the specific realization of the $u_i^\alpha$
and $v_j^\alpha$ is considered to be irrelevant, and can be averaged out with respect to some simple distributions, as long as 
this average is correlated with the data. This means that the decomposition $\hat s_\alpha = \sum_i u_i^\alpha \hat s_i$ of any given sample configuration is
assumed also to be kept fixed while averaging. What matters mainly is the strength 
given by $w_\alpha(t)$ and the rotation given by $\Omega_{\alpha\beta}^{v,h}(t)$ of these SVD modes.  
Assuming for example i.i.d centered normal distribution with respective variance $1/N_v$ and $1/N_h$ for $u_i^\alpha$
and $v_j^\alpha$, the empirical term takes the simple form:

\begin{strip}
\begin{equation}\label{eq:empterm}
\langle s_\alpha\sigma_\beta\rangle_{\rm Data}  = \frac{1}{N_h}\Bigl\langle s_\alpha(s_\beta w_\beta-\theta_\beta)V\Bigl(\frac{1}{N_h}\sum_\gamma (w_\gamma s_\gamma-\theta_\gamma)^2\Bigr)\Bigr\rangle_{\rm Data}
\ 
\text{where}\ 
V(x) = \int dy\frac{e^{-y^2/2}}{\sqrt{2\pi}}\sech^2(\sqrt{x}y),
\end{equation}
\end{strip}

which actually  depends on the activation function (an hyperbolic tangent in this case). The main point here is that the empirical term defines an operator
whose decomposition onto the SVD modes of $W$ functionally depends solely on $w_\alpha,\theta_\alpha$ and on the projection of the data on the SVD modes of $W$.
This term is precisely driving the dynamics. The adaptation of the RBM to this driving force is given by the second term which can be as well 
estimated in the thermodynamic limit, as a function of $w_\alpha$, $\theta_\alpha$ and $\eta_\alpha$ alone. 

\section{Linear instabilities}\label{sec:linear}
\begin{figure}
	\centering
	\includegraphics[scale=0.45]{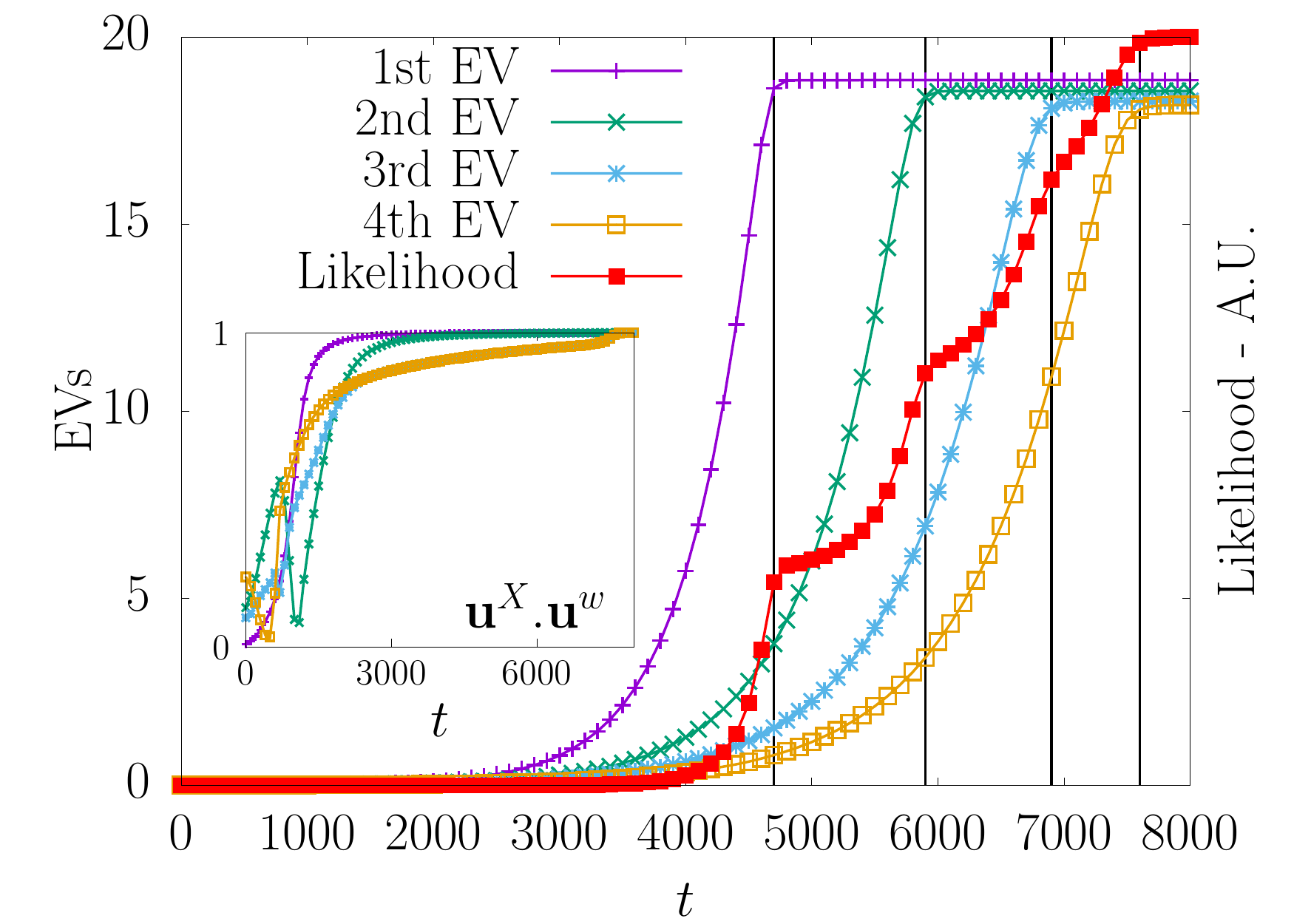}
	\caption{Time evolution of the eigenvalues in the linear model and of the likelihood. We observe very clearly how the different modes emerge from the bulk and how the likelihood increases at each eigenvalue learned. In the inset, the scalar product of the vectors $\bm{u}$ obtained from the SVD of the data and of $\bm{w}$. The $\bm{u}$s of $\bm{w}$ are aligned with the SVD of the data at the end of the learning.}
	\label{fig_linRBM_dyn}
\end{figure}

First let us consider the linear regime which can be analyzed thoroughly. 
It can be obtained by rescaling all the weights and fields by a common ``inverse temperature'' $\beta$ factor and let this go to zero 
in equations~(\ref{eq:cd1}). This limit can be understood by keeping up to quadratic terms in the mean field free energy and should correspond
to the first stages of the learning. In this limit, magnetizations $(\mu_v,\mu_h)$ of visible and hidden variables have Gaussian fluctuations with covariance matrix
\begin{equation*}
C(\mu_v,\mu_h) \egaldef 
\left[
\begin{matrix}
\sigma_v^{-2} & - W \\[0.2cm]
-W^T & \sigma_h^{-2} 
\end{matrix}
\right]^{-1}
\end{equation*}
with $\sigma_v^2 = \sigma_h^2 = 1$ introduced for sake of generality when considering general linear RBM.
To simplify the exposition,  we discard the biases of the data and related fields ($\theta_\alpha,\eta_\alpha)$ of the RBM. 
In that case the empirical term in (\ref{eq:cd1}) involves directly the 
covariance matrix of the data expressed in the frame defined by the SVD modes of $W$ 
\begin{equation*}
\langle s_\alpha\sigma_\beta\rangle_{\rm Data} = \sigma_h^2 w_\beta \langle s_\alpha s_\beta\rangle_{\rm Data}. 
\end{equation*}
From $C(\mu_v,\mu_h)$ we get the other terms yielding the following equations:
\begin{align*}
  \frac{dw_\alpha}{dt} &= w_\alpha\sigma_h^2\Bigl(\langle s_\alpha^2\rangle_{\rm Data} - \frac{\sigma_v^2}{1-\sigma_v^2\sigma_h^2w_\alpha^2}\Bigr) \\
  \Omega_{\alpha\beta}^{v,h} &=  
(1-\delta_{\alpha\beta})\sigma_h^2\Bigl(\frac{w_\beta-w_\alpha}{w_\alpha+w_\beta}\mp\frac{w_\beta+w_\alpha}{w_\alpha-w_\beta}\Bigr)\langle s_\alpha s_\beta\rangle_{\rm Data}
\end{align*}
Note that these equations are exact for a linear RBM, since they can be derived without any reference to the coordinates of $u_\alpha$ and $v_\alpha$
over which we average in the non-linear regime. These equations tell us that, during the learning the vectors 
$\bm{u}^\alpha$ (and also $\bm{v}^\alpha$) will rotate until being aligned to the the principal components of the data, 
i.e. until $\langle s_\alpha s_\beta\rangle_{\rm Data}$ becomes diagonal.
Then calling $\hat w_\alpha^2$ the corresponding empirical variance given by the data, 
the system reach the following equilibrium values:
\begin{equation*}
w_\alpha^2 = 
\begin{cases}
\DD \frac{\hat w_\alpha^2-\sigma_v^2}{\sigma_v^2\sigma_h^2 \hat w_\alpha^2}\qquad\ \ \text{if} \qquad \hat w_\alpha^2 > \sigma_v^2,\\
\DD 0\qquad\qquad\qquad \text{if} \qquad \hat w_\alpha^2 \le \sigma_v^2.
\end{cases}
\end{equation*}
From this we see that the RBM selects the strongest SVD modes in the data.
The linear instabilities correspond to directions for which the variance of the data is above the threshold $\sigma_v^2$.
This determines the deformations of the weight matrix which can develop during the learning and will eventually interact, following the usual
mechanism of non-linear pattern formation like e.g. in reaction-diffusion processes~\cite{Cross-Hohenberg}. 
Other possible deformations are damped to zero. The linear RBM will therefore learn all (up to $N_h$) principal 
components that passed the threshold but it is important to remember that the resulting distribution will still be unimodal. 
Note that this selection mechanism is already known to occur for linear auto-encoders~\cite{Bou-Kamp} 
or some other similar linear Boltzmann machines~\cite{TiBi}.
On Fig.~\ref{fig_linRBM_dyn} we can see the eigenvalues being learned one by one in a linear RBM. For non-linear RBM when the system escapes the linear regime, 
a well suited mean-field theory is required to understand the dynamics and the steady-state regime.
 
\section{Non-linear regime}
During the linear regime some specific modes are selected and at some point these modes start to interact in a non-trivial manner. The empirical 
term in~(\ref{eq:cd1}) involves higher order statistics of the data as exemplified by~(\ref{eq:empterm}) and the Gaussian estimation with $\sigma_v^2=\sigma_h^2=1$ 
of the RBM response term $\langle s_\alpha\sigma_\beta\rangle_{\rm RBM}$ is no longer valid. 
In order to estimate this term in the thermodynamic limit, some assumptions on the form of the 
weight matrix are needed. A common assumption consists in considering i.i.d. random variables for the weights $w_{ij}$ and this, like for example in~\cite{TuMo,Barra,HHuang}, generally leads to a Marchenko-Pastur distribution of the singular values of $W$, 
which as we shall see in the next section is unrealistic. Instead, based on our experiments such distribution corresponds to the noise of the weight matrix, 
while its information content is better expressed by the presence of SVD modes outside of the bulk.
This leads us to write the weight matrix as  
\begin{equation}\label{eq:wsvd}
	w_{ij} = \sum_{\alpha=1}^{K} w_\alpha u_i^{\alpha} v_j^{\alpha} + r_{ij}
\end{equation}
where the $w_\alpha = O(1)$ are isolated singular values (describing a rank $K$ matrix), the $\bm{u}^\alpha$ and $\bm{v}^\alpha$ are the eigenvectors of the SVD 
decomposition and the $r_{ij}={\cal N}(0,\sigma^2/L)$ where $L=\sqrt{N_h N_v}$ are i.i.d. corresponding to noise.  To be consistent with the linear analysis, 
these modes are assumed to span the (left) subspace corresponding to the part of the empirical SVD above threshold while $r$ spans the complementary space 
of empirical modes below threshold.
We limit the analysis here to the case where $K$ is finite.
This then allows us to assume simple distributions $p_u$ and $p_v$ for the components of $\bm{u}^\alpha$ and $\bm{v}^\alpha$ considered i.i.d. for instance.
This altogether defines our statistical ensemble of RBM to which we restrict ourselves to study the learning procedure. 
For $K$ extensive we should instead average over the orthogonal group which would lead to a slightly 
different mean-field theory~\cite{PaPo,OpWi}. 
In the present form our model of RBM is similar 
to the Hopfield model and recent generalizations~\cite{Mezard}, 
the patterns being represented by the SVD modes outside the bulk. The main difference, in addition to the bipartite structure of the graph, 
is the non-degeneracy of the singular values $w_\alpha$. Still the analysis in the thermodynamic limit follows classical treatments 
like~\cite{AmGuSo1,AmGuSo3} for the Hopfield model or~\cite{Barra} for bipartite models.
The starting point is to express the average over $u,v$  and weights $r_{ij}$ of the log partition function $Z$ in~(\ref{eq_proba_rbm}) with the help of the replica trick:
\begin{equation*}
\EE_{u,v,r}[\log(Z)] = \lim_{p\to 0}\frac{d}{dp}\EE_{u,v,r}[Z^p].\\[0.2cm]
\end{equation*}
After averaging over the iid weights, 4 sets of order parameters $\{(m_\alpha^a,\bar m_\alpha^a),a=1,\ldots p,\alpha=1,\ldots K\}$ 
and $\{(Q_{ab},\bar Q_{ab}),a,b=1,\ldots p, a\ne b\}$  are introduced with help of two distinct 
Hubbard-Stratonovich transformations. These variables represent the following quantities:
\begin{align*}
m_\alpha^a \sim \frac{1}{\sqrt{L}} E_{u,v,r}\bigl(\langle \sigma_\alpha^a\rangle\bigr)\qquad
\bar m_\alpha^a \sim \frac{1}{\sqrt{L}} E_{u,v,r}\bigl(\langle s_\alpha^a\rangle\bigr)\\[0.2cm]
Q_{ab} \sim E_{u,v,r}\bigl(\langle \sigma_i^a \sigma_i^b\rangle\bigr)\qquad
\bar Q_{ab} \sim E_{u,v,r}\bigl(\langle s_j^a s_j^b\rangle\bigr),
\end{align*}
namely the correlations of the hidden [resp. visible] states with the left [resp. right] 
singular vectors and the Edward-Anderson order parameters measuring the correlation
between replicas of hidden or visible states. 
$\EE_{u}$ and  $\EE_{v}$ denote an average  wrt to the rescaled components $u\simeq \sqrt{N_v}u_i^\alpha$ and $v\simeq \sqrt{N_h}v_j^\alpha$ of the SVD modes.
The transformations involve pairs of complex integration variables because of the asymmetry introduced by the two-layers structure 
by contrast to fully connected models.
They lead to the following representation:

\begin{strip}
\begin{align*}
\EE_{u,v,r}[Z^p] &= \int \prod_{a,\alpha}\frac{dm_\alpha^a d\bar m_\alpha^a}{2\pi}\prod_{a\ne b}\frac{dQ_{ab}d\bar Q_{ab}}{2\pi}
\exp\Bigl\{-L\Bigl(\sum_{a,\alpha}w_\alpha m_\alpha\bar m_\alpha+\frac{\sigma^2}{2}\sum_{a\ne b}Q_{ab}\bar Q_{ab}-\frac{1}{\sqrt{\kappa}}A[m,Q]
-\sqrt{\kappa}B[\bar m,\bar Q]\Bigr)\Bigr\}\\[0.2cm]
\text{with}&\qquad
A[m,Q] \egaldef \log\Bigl[\sum_{S^a\in\{-1,1\}}\EE_u \Bigl(e^{\frac{\sqrt{\kappa}\sigma^2}{2}\sum_{a\ne b}Q_{ab}S^a S^b +\kappa^{\frac{1}{4}}\sum_{a,\alpha}(m_\alpha^a w_\alpha-\eta_\alpha)u^\alpha S^a}\Bigr)\Bigr],
\end{align*}
\end{strip}

$\kappa=N_h/N_v$
and $B[\bar m,\bar Q]$ obtained from $A[m,Q]$ by replacing $u$ by $v$, $\eta$ by $\theta$ and $\kappa$ by $1/\kappa$.
The thermodynamic properties are obtained by first letting $L\to\infty$ allowing for a saddle point approximation and then the limit $p\to 0$ is taken.
We restrict here the discussion to replica symmetric (RS) saddle points~\cite{MePaVi}. The breakdown of RS can actually be determined  
by computing the so-called AT line~\cite{AlTh} and will be detailed somewhere else~\cite{DeFiFu}. 
In the RS case the set $\{(Q_{ab},\bar Q_{ab}\}$ reduces to a pair $(q,\bar q)$  of spin glass parameters,  
while  quenched magnetization towards the SVD directions are now represented by
$\{(m_\alpha,\bar m_\alpha),\alpha=1,\ldots K\}$.
Letting $x={\cal N}(0,1)$ and skipping some details, the saddle-point equations are given by 
\begin{align}
(m_\alpha,\bar m_\alpha) &= \EE\Bigl(\kappa^{\frac{1}{4}}v^\alpha\tanh\bigl(\bar h(x,v)\bigr),
\kappa^{-\frac{1}{4}}u^\alpha\tanh\bigl(h(x,u)\bigr)\Bigr)\label{eq:mf1}\\[0.2cm]
(q,\bar q) &= \EE\Bigl(\tanh^2\bigl(\bar h(x,v)\bigr),\tanh^2\bigl(h(x,u)\bigr)\Bigr),\label{eq:mf2}
\end{align}
with $\EE$ denoting the average over $(u,v,x)$ and 
\begin{align*}
h(x,u) &\egaldef  \kappa^{\frac{1}{4}}\bigl(\sigma\sqrt{q}x+\sum_\gamma (w_\gamma m_\gamma-\eta_\gamma) u^\gamma\bigr)\\[0.2cm]
\bar h(x,v) &\egaldef \kappa^{-\frac{1}{4}}\bigl(\sigma\sqrt{\bar q}x+\sum_\gamma (w_\gamma\bar m_\gamma-\theta_\gamma) v^\gamma\bigr).
\end{align*}

These fixed point equations can be solved numerically to tell us how the variables condensate on the SVD modes within each equilibrium state of the distribution
and whether a spin glass phase is present or not. The important point here is that 
with $K$ finite and a non-degenerate spectrum the mode with highest singular value dominates the ferromagnetic phase. 
The phase diagram looks in fact similar to the one of the SK model with ferromagnetic coupling, when $1/\sigma$ is interpreted as a temperature
and $w_{max}/\sigma$ the ferromagnetic coupling. 
Some subtleties arise when considering various ways of averaging over singular vectors components~\cite{DeFiFu}. 
In~\cite{Agliari,TuMo} it is underlined the importance of the capability of networks to produce  compositional states 
structured by combination of hidden variables. In our representation, we don't have direct access to this property, 
but to the dual one in some sense, namely states corresponding to combination of modes.
Their presence and their structure, are rather sensitive to the way the average over $u$ and $v$ is performed. 
In this respect the case where $\bm{u}^\alpha$ and $\bm{v}^\alpha$ are Gaussian i.i.d distributed is very special: 
all other fixed points associated to lower modes can be shown to be unstable as well as fixed points associated to combinations of modes. 
Instead, for other distributions with smaller kurtosis, like uniform or Bernoulli, stable fixed points associated to many different single modes or 
combinations of modes can exist and contribute to the thermodynamics.

Coming back to the learning dynamics, the first thing which is expected, already from the linear analysis, is that the noise term in (\ref{eq:wsvd}) vanishes
by condensing into a delta function of zero modes. 
Then the term corresponding to the response of the RBM in~(\ref{eq:cd1}) is estimated (in absence of bias) in the thermodynamic by means of 
the order parameters defined previously:
\begin{equation*}
\langle s_\alpha\sigma_\beta\rangle_{\rm RBM} = \frac{L}{Z_\text{\tiny MF}}\sum_{q=1}^C e^{-F_q}\bar m_\alpha^{(q)} m_\beta^{(q)},\ \  
Z_{MF} \egaldef \sum_q e^{-F_q}
\end{equation*}
where the index $q$ run over all stable fixed point solutions of~(\ref{eq:mf1},\ref{eq:mf2}) weighted accordingly to their free energy.
These are the dominant contributions as long as free energy differences are $O(1)$, 
internal fluctuations given by each fixed point are comparatively of order $O(1/L)$. 
Note that this is the reason why the RBM needs to reach a 
ferromagnetic phase with many states to be able to match the empirical term in~(\ref{eq:cd1}) in order to converge.
For instance, in the case of a multimodal data distribution with many well separated clusters, 
the SVD modes of $W$ which will develop are the one pointing in the direction of the magnetizations defined by these clusters. 
In this simple case the RBM will evolve as in the linear case to a state such that the empirical term  becomes diagonal, while the singular values 
adjust themselves until matching the proper magnetization in each fixed point.  
More precise statements about the phase diagram of the RBM  and the behaviour of our dynamical equations 
including the dynamics of the external fields $\eta_\alpha$ and $\theta_\alpha$ will be given in~\cite{DeFiFu}. 

\section{Tests on the MNIST dataset}

\begin{figure}
  \centering
  \begin{subfigure}{.32\linewidth}
    \includegraphics[width=\linewidth,trim=30 0 50 0,clip]{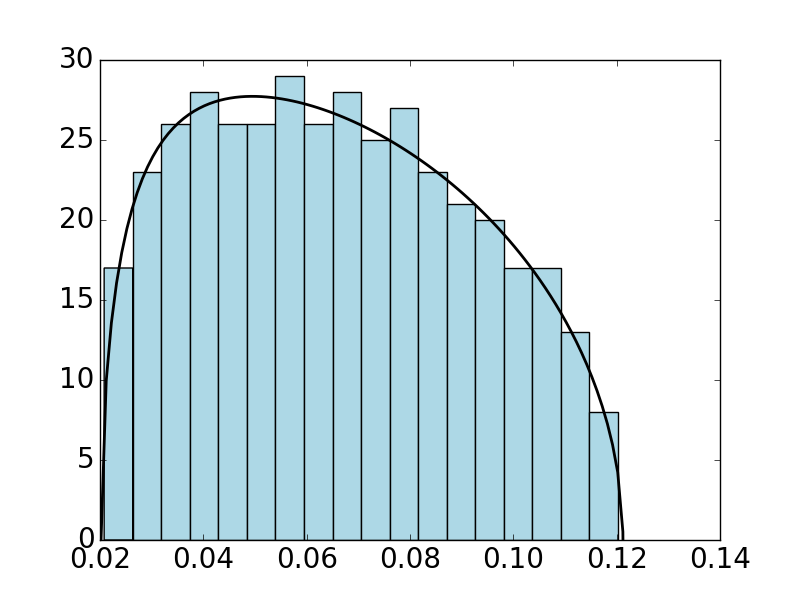}
    \caption{}
    \label{fig:mp_fit}
  \end{subfigure}
  \begin{subfigure}{.32\linewidth}
    \includegraphics[width=\linewidth,trim=30 0 50 0,clip]{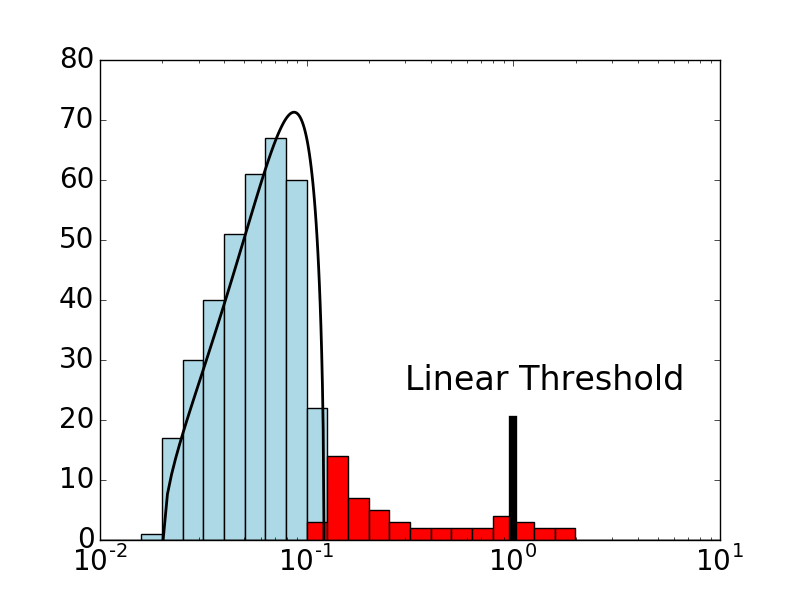}
    \caption{}
    \label{fig:sv1}
  \end{subfigure}
  \begin{subfigure}{.32\linewidth}
    \includegraphics[width=\linewidth,trim=30 0 50 0,clip]{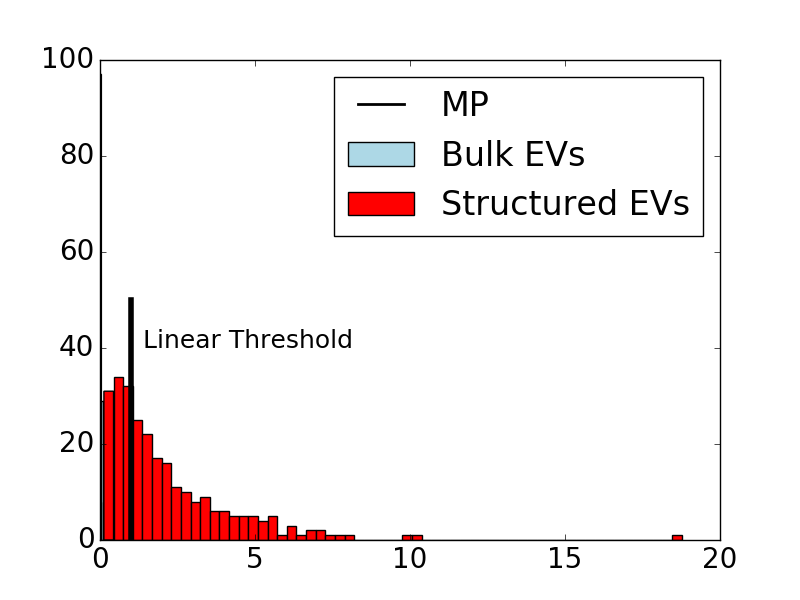}
    \caption{}
    \label{fig:sv2}
  \end{subfigure}\par\medskip
 \caption{\textbf{(a)} Singular values distribution of the initial random matrix compared to Marchenko-Pastur law. \textbf{(b)} With the training we can see some singular values strengthening and overcoming the threshold set by the Marchenko-Pastur law. \textbf{(c)} Distribution of the singular values after a long training: we can see many outliers spread above threshold and a spike of below-threshold singular values near zero.}
 \label{fig_mnist_svd_modes}
\end{figure}

We illustrate our results on the MNIST dataset. The MNIST dataset is composed of $60000$ images of handwritten digits of $28 \times 28$ pixels. It is known that RBMs perform reasonably well on this dataset and therefore we can now interprete in the light of the preceding sections how the learning goes. For the training of the MNIST dataset we use the following parameters.  The weights of the matrix $W$ were initiated randomly from a centered Gaussian distribution with a variance of $0.01$ such that the MP bulk do not pass the threshold. The visible fields are initialized to reproduce the empirical mean of the data for each visible variable. The hidden field is put to zero. The learning rate is chosen to be $\approx 0.01$. With these parameters we verified that our machine was able to sample digits in a satisfactory way after $20$ epochs. Now we can investigate the value of some observables introduced previously. First, we look at the SVD modes of the matrix $w$ during the learning on Fig.~\ref{fig_mnist_svd_modes}. We see that, after seeing only few updates the system has already learned many SVD modes from the data.

%

\begin{figure}
  \centering
  \begin{subfigure}{.15\linewidth}
    \includegraphics[width=\linewidth]{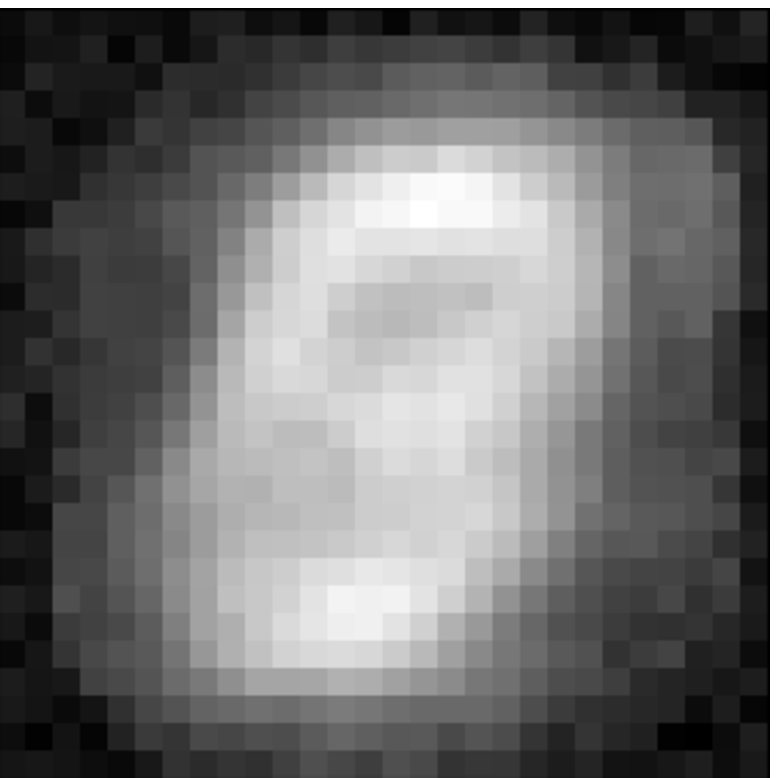}
    \caption{}
    \label{fig:m1tr}
  \end{subfigure}
  \begin{subfigure}{.15\linewidth}
    \includegraphics[width=\linewidth]{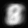}
    \caption{}
    \label{fig:vbias}
  \end{subfigure}
  \begin{subfigure}{.15\linewidth}
    \includegraphics[width=\linewidth]{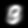}
    \caption{}
    \label{fig:m1data}
  \end{subfigure}\par\medskip
  \begin{subfigure}{\linewidth}
    \centering
    \includegraphics[width=.08\linewidth]{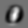}
    \includegraphics[width=.08\linewidth]{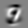}
    \includegraphics[width=.08\linewidth]{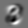}
    \includegraphics[width=.08\linewidth]{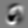}
    \includegraphics[width=.08\linewidth]{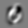}
    \includegraphics[width=.08\linewidth]{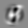}
    \includegraphics[width=.08\linewidth]{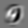}
    \includegraphics[width=.08\linewidth]{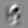}
    \includegraphics[width=.08\linewidth]{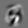}
    \includegraphics[width=.08\linewidth]{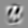}
    \caption{}
    \label{fig:modes_data}
  \end{subfigure}
  \begin{subfigure}{\linewidth}
  	\centering
    \includegraphics[width=.08\linewidth]{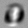}
    \includegraphics[width=.08\linewidth]{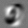}
    \includegraphics[width=.08\linewidth]{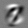}
    \includegraphics[width=.08\linewidth]{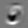}
    \includegraphics[width=.08\linewidth]{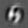}
    \includegraphics[width=.08\linewidth]{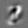}
    \includegraphics[width=.08\linewidth]{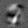}
    \includegraphics[width=.08\linewidth]{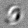}
    \includegraphics[width=.08\linewidth]{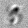}
    \includegraphics[width=.08\linewidth]{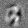}
    \caption{}
    \label{fig:modes_tr}
  \end{subfigure}\par
  \begin{subfigure}{\linewidth}
    \centering
    \includegraphics[width=.08\linewidth]{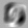}
    \includegraphics[width=.08\linewidth]{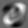}
    \includegraphics[width=.08\linewidth]{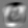}
    \includegraphics[width=.08\linewidth]{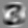}
    \includegraphics[width=.08\linewidth]{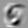}
    \includegraphics[width=.08\linewidth]{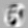}
    \includegraphics[width=.08\linewidth]{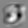}
    \includegraphics[width=.08\linewidth]{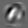}
    \includegraphics[width=.08\linewidth]{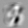}
    \includegraphics[width=.08\linewidth]{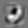}
    \caption{}
    \label{fig:modes_tr_10}
  \end{subfigure}
   \caption{\textbf{(a)} First mode learnt by the RBM with the external visible field initialized as a null vector. \textbf{(b)} External visible field initialized on the empirical mean. \textbf{(c)} First principal components extracted from the training set. \textbf{(d)} Principal components extracted from the training set (starting from the second). \textbf{(e)} The first 10 modes of a RBM trained for 1 epoch. \textbf{(f)} Same as (e) but after a 10 epochs training.}
  \label{fig_plot_svdmode}
\end{figure}

On Figure~\ref{fig:mp_fit}-\ref{fig:sv2}, we observe what is expected from the linear regime. Some modes escape from the Marchenko-Pastur bulk of the eigenvalues while other condense down to zero. In particular, we can see that the modes at the beginning of the learning correspond exactly to the SVD modes of the data, see Fig.~\ref{fig_plot_svdmode}. On this figure, we notice that the modes of the $W$ matrix are the same as the ones of the data at the beginning of the learning as predicted by the linear theory.

\begin{figure}
  \includegraphics[width=0.9\linewidth]{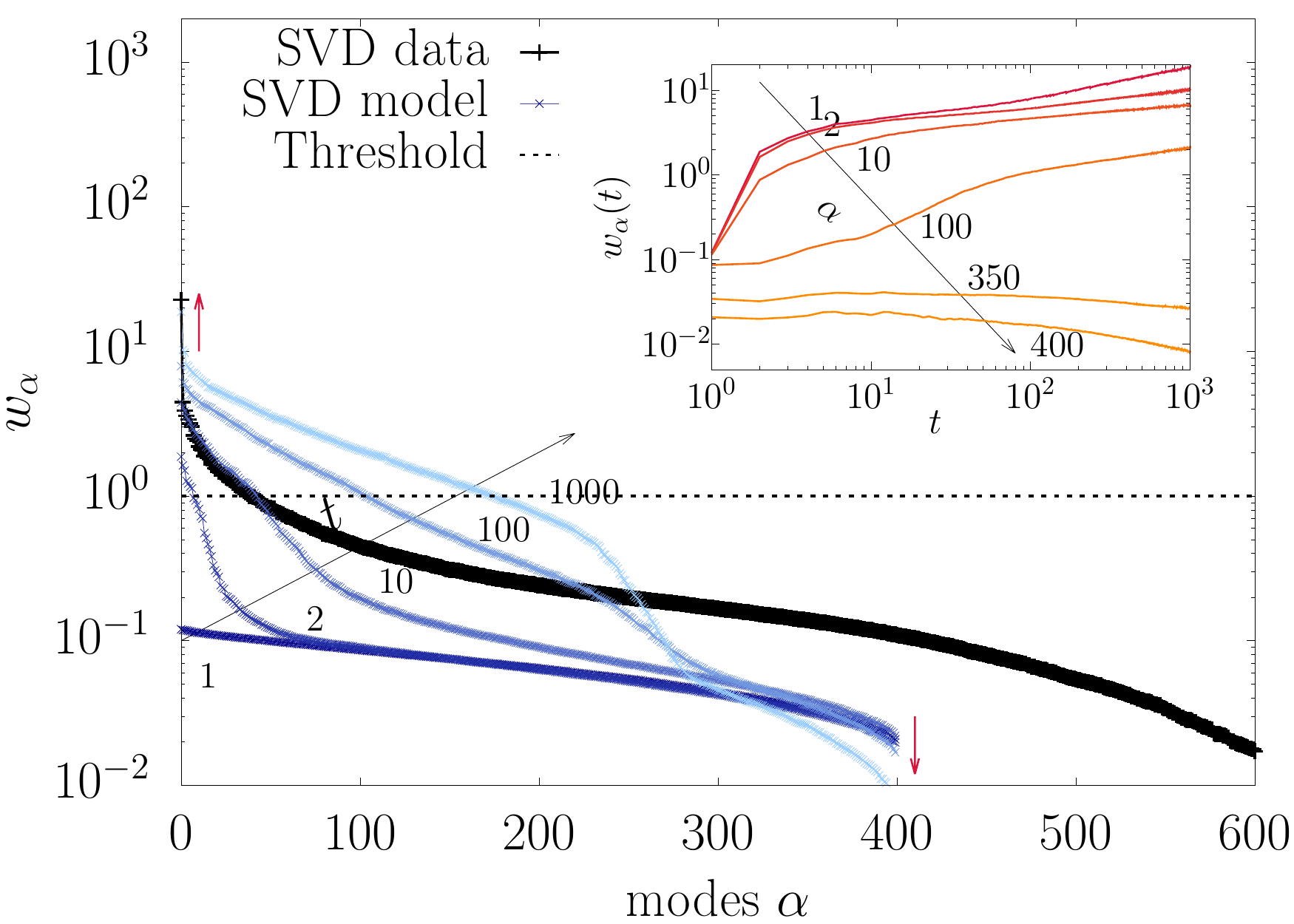}
  \caption{Log-log plot of the singular values represented as discrete abscissas (in decreasing order) with their magnitude reported on the ordinates. The RBM contained $400$ hidden variables. A cutoff is highlighted by the onset of the linear behaviour and the SVD modes of the data in black. We qualitatively observe that beyond some $\alpha_{\rm tresh}$ the modes are dumped while before they are amplified. In the inset, the time evolution of the modes 1, 2, 10, 100, 350, 400 during the learning as a function of the number of epochs, we see that for large	 value of $\alpha$, the modes are decreasing. We observe that the linear cutoff (around $\alpha \approx 50$ seems different from the one observed when going deep into the non-linear regime ($\alpha \approx 250$).}
  \label{fig:sva_pl}
\end{figure}

After many epochs, we observe on  Fig.~\ref{fig_plot_svdmode}-f that non-linear effects have deformed the SVD modes of $W$ with respect to the beginning of the learning. We can also look at the evolution of the eigenvalues of $W$. On Fig.~\ref{fig:sva_pl} we observe their evolution and when they start to be amplified (or dumped). On the inset, we see how the strongest mode get out of the bulk and increase while the lowest ones are dumped after many epochs. We also observe that the top part of the spectrum of $W$ appear flattened as compared to empirical SVD spectrum. This presumably favors the expression of many states of similar free energy related to various digit configurations, able to contribute to RBM response term in~(\ref{eq:cd1}).

\section{Discussion}

The equations obtained for the dynamics and the MF theory that allows us to compute them 
constitute a phenomenological description of the learning of an RBM. This is assumed to 
represent a typical learning trajectory in the limit of infinite batch size. These equations have been obtained by averaging over 
the components of left and right SVD vectors of the weight matrix, keeping fixed a certain number of quantities considered to 
be the relevant ones, fully characterizing a typical RBM during the learning process. This averaging 
corresponds actually  to a standard self-averaging assumption in a RS phase.
The singular values spectrum $\{w_\alpha\}$ is playing the main role. The projections $(\eta_\alpha,\theta_\alpha)$ of the bias onto the eigenmodes 
of $W$ are also considered as intrinsic quantities. Finally the rotation vectors $\{\Omega_{\alpha,\beta}^{v,h}\}$
give us the relative motion of the data w.r.t the time dependent frame given by the singular vectors of $W$.
In our phenomenological description the learning dynamics is 
represented by a trajectory of $\{w_\alpha(t),\eta_\alpha(t),\theta_\alpha(t),\Omega_{\alpha\beta}^{v,h}(t)\}$ which is uniquely determined by our 
equations once an initial condition specified by the decomposition of the data on the singular vectors of $W$ is given.
By contrast to usual approaches which rely on the teacher-student scenario, 
we may obtain generic learning curves of non-linear neural networks, which are driven by intrinsic properties of the data.
The point is to give insights into the relationship between model and data. This allows us to give some elements of understanding on 
which properties of the data drive the learning and how they are represented in the model. Eventually this will lead us to identify and cure 
some flaws of present learning methods.

\acknowledgments
We thank B. Seoane helping us improving the quality of the letter.

\bibliographystyle{eplbib}

\bibliography{rbm}

\end{document}